\documentclass[nofootinbib,aps,pra,showpacs,superscriptaddress,preprint]{revtex4}
\usepackage{amsmath}
\usepackage{subfig}
\usepackage{graphicx}

\catcode`ð=\active
\defð{\u{g}}
\catcode`Ð=\active
\defÐ{\u{G}}
\catcode`Ý=\active
\defÝ{\. I}
\catcode`ö=\active
\defö{\"{o}}
\catcode`Ö=\active
\defÖ{\"O}
\catcode`ü=\active
\defü{\"{u}}
\catcode`Ü=\active
\defÜ{\"{U}}
\catcode`Þ=\active
\defÞ{\c{S}}
\catcode`þ=\active
\defþ{\c{s}}
\catcode`ý=\active
\defý{{\i}}
\catcode`ç=\active
\defç{\d{c}}
\catcode`Ç=\active
\defÇ{\d{C}}

\begin{document}

\title{Scattering of a spinless particle by an asymmetric Hulth{\'e}n potential within the effective mass formalism}
\author{\small Oktay Aydo\u{g}du}
\email[E-mail: ]{oktaydogdu@gmail.com}\affiliation{Department of
Physics, Mersin University, 33343, Turkey}
\author{\small Altuð Arda}
\email[E-mail: ]{arda@hacettepe.edu.tr}\affiliation{Department of
Physics Education, Hacettepe University, 06800, Ankara,Turkey}
\author{\small Ramazan Sever}
\email[E-mail: ]{sever@metu.edu.tr}\affiliation{Department of
Physics, Middle East Technical  University, 06531, Ankara,Turkey}

\begin{abstract}
Effective mass Klein-Gordon equation for the asymmetric
Hulth{\'e}n potential is solved in terms of hypergeometric
functions. Results are obtained for the scattering and bound
states with the position dependent mass and constant mass, as a
special case. In both cases, we derive a condition for the
existence of transmission resonance ($T=1$). We also study how the
transmission resonance
depends on the particle energy and the shape of the external potential.
\end{abstract}
\date{\today}

\maketitle

\newpage

\section{Introduction}
Solution of the Schr{\"o}dinger equation for an external potential
for the bound and scattering states [1] is a fundamental problem.
In the low-momentum limit $k \rightarrow 0$ $(E\rightarrow 0)$,
the transmission and reflection coefficients for a physical
potential are well behaved at infinity in one dimension going to
$0$ and $1$, respectively, unless the external potential supports
a bound state for this limit [2]. In that case, the zero energy
resonance (half-bound state) described by the not square
integrable wave function is finite at infinity [1, 3]. So
transmission coefficient goes to $1~(unity)$ while reflection
coefficient goes to $0$. This phenomenon is called as
\textit{transmission resonance} [4]. On the other hand, a
condition for the existence of the transmission resonance in view
of asymmetric potentials has been recently investigated in Refs.
[5, 6] for the non-relativistic particles. The transmission
resonance concept has been recently generalized to the
relativistic case [7, 8, 9, 10]. Dombey and Kennedy [8] showed
that in the low-momentum limit $k \rightarrow 0$, Dirac particles
scattered by an external potential have half-bound states at
$E=\pm m$  in contrast to non-relativistic particles where
half-bound state occurs only at zero energy. Thus, we should speak
of zero momentum resonances in the relativistic case rather than
zero energy resonances [8]. Afterwards, the scattering and bound
state solutions of the Dirac equation for the Woods-Saxon
potential have been obtained in the low-momentum limit. Conditions
for a zero momentum resonance (transmission resonance: $T = 1$)
and supercriticality (as the particle bound state at $E=-m$) have
been derived in Ref. [9]. After these pioneering studies, the
transmission resonance and supercriticality for the
relativistic/non-relativistic particles in an external potentials
have been extensively discussed [3, 11, 12, 13, 14, 15, 16, 17,
18]. In one of these studies [13], the authors showed that the
transmission coefficient obtained for the Klein-Gordon particle
displays a behavior similar to that of the one obtained for the
Dirac particle [8].

Recently, solving the non-relativistic/relativistic wave equations
with external potential and obtaining the bound states have been
widely studied in view of the position-dependent mass formalism.
It has extensive applications in condensed matter physics and
material science such as electronic properties of the
semi-conductors [19], quantum dots [20], and quantum liquids [21].
Besides, the scattering problem in the position-dependent mass
framework has been recently received much attentions [22, 23, 24,
25, 26] . The effective mass Dirac equation for the Coulomb field
has been investigated in Ref. [22]. Dutra~{\it et al.} have
obtained exact solution of the Dirac equation for the inversely
linear potential in the presence of the position-dependent mass
[24].

In the present work, we investigate the transmission resonances
for the Klein-Gordon particle scattered by the asymmetric
Hulth{\'e}n potential within the position-dependent mass
formalism. The Hulth{\'e}n potential [27] is one of the most
significant short-range potential in physics and has been used in
atomic physics, condensed matter, nuclear and particle physics,
and chemical physics [28, 29, 30, 31]. The generalized Hulth{\'e}n
potential is a potential containing several potential forms such
as usual Hulth{\'e}n potential, Woods-Saxon potential, Cusp
potential and Coulomb potential. Recently, Sogut [32] have
obtained the exact solution of the one-dimensional
Duffin-Kemmer-Petiau equation for the asymmetric Hulth{\'e}n
potential and investigated the bound and scattering states of
vector bosons.

The asymmetric Hulth{\'e}n potential is given in the following
form [32]
\begin{equation}
V_{AHP}=V_0\left[\frac{\Theta(-x)}{e^{-a
x}-q}+\frac{\Theta(x)}{e^{b x}-\tilde{q}}\right]\,,\label{1}
\end{equation}
where $V_0$ is the strength of the potential, $a$, $b$, $q~(<1)$
and $\tilde{q}~(<1)$ are the positive parameters related to the
shape of potential. $\Theta(x)$ is the Heaviside step function. It
is worth talking about that the asymmetric Hulth{\'e}n potential
transforms to the usual Hulth{\'e}n potential for $a=b$ and
$q=\tilde{q}$ and the asymmetric Cusp potential for
$q=\tilde{q}=0$. The form of the asymmetric Hulth{\'e}n potential
is shown in Fig. 1.

We take mass distribution as
\begin{equation}
m(x)=m_0+m_1\,f(x)\,,\label{101}
\end{equation}
where $m_0$ and $m_1$ are positive parameters and the function is
given as $f(x)=\frac{\Theta(-x)}{e^{-a x}-q}+\frac{\Theta(x)}{e^{b
x}-\tilde{q}}$. This form of the mass function makes it possible
to solve the problem analytically.

The paper is organized as follows: In the next section, we find
the exact solution of the Klein-Gordon equation in terms of
hypergeometric functions. In section 3, transmission and
reflection coefficients are obtained by using asymptotic behavior
of the hypergeometric functions. In section 4, we investigate the
bound states. Section 5 is devoted to discussions. Finally, we
summarize the results in the last section.

\section{Effective mass Klein-Gordon equation for the asymmetric Hulth{\'e}n potential}
In $( 1 + 1 )$ dimensions, the time-independent Klein-Gordon
equation with scalar $S(x)$ and vector $V(x)$ potentials in the
presence of the effective  mass can be written as [24]
\begin{equation}
\frac{d^2\Psi(x)}{d
x^2}+\left\{\left[E-V(x)\right]^2-\left[m(x)+S(x)\right]^2\right\}\Psi(x)=0\label{2}
\end{equation}
where $E$ is the energy of the relativistic particle. Here, we
take $\hbar=c=1$ for the simplicity. We investigate the scattering
and bound state solutions of the Eq.~(\ref{2}) by using the mass
distribution given in Eq.~(\ref{101}) together with the following
scalar $S(x)$ and vector $V(x)$ potentials
\begin{eqnarray}
S(x)&=&S_0\,f(x)\label{3}\,,\\
V(x)&=&V_0\,f(x)\label{4}\,,
\end{eqnarray}
where $S_0$ and $V_0$ are positive parameters.

In order to find the scattering of a Klein-Gordon particle from
the asymmetric Hult{\'e}n potential in the presence of the
effective mass, we first seek the solution of the Klein-Gordon
eqution for $x<0$. In that case, Eq.~(\ref{2}) becomes
\begin{equation}
\frac{d^2\Psi_L(x)}{d
x^2}+\left\{\left[E-\frac{V_0}{e^{-ax}-q}\right]^2-\left[m_0+\frac{m_1+S_0}{e^{-ax}-q}\right]^2\right\}\Psi_L(x)=0\label{6}.
\end{equation}
Changing the variable $e^{-ax}=q/y$, Eq.~(\ref{6}) takes the
following form
\begin{eqnarray}
y(1-y)^2\left(y\frac{d^2\Psi_L(y)}{d y^2}+\frac{d\Psi_L(y)}{d
y}\right)
+\left\{\left[\left(E+\frac{V_0}{q}\right)^2 -\left(m_0-\frac{m_1+S_0}{q}\right)^2\right]\right.\nonumber\\
\left.\times\frac{y^2}{a^2}-2\left(E^2-m_0^2+\frac{E
V_0}{q}+\frac{m_0(m_1+S_0)}{q}\right)\frac{y}{a^2}+\frac{(E^2-m_0^2)}{a^2}\right\}
\Psi_L(y)=0\label{7}.
\end{eqnarray}
By setting $\Psi_{L}(y)=y^{\mu}(1-y)^{\nu}H(y)$ and substituting
it into the above equation, one gets the hypergeometric equation
[33]
\begin{equation}
y(1-y)\frac{d^2H(y)}{dy^2}+[2\mu+1-(2\mu+2\nu+1)y]\frac{dH(y)}{dy}-(\mu+\nu-\gamma)(\mu+\nu+\gamma)H(y)=0\label{8}
\end{equation}
where
\begin{eqnarray}
\mu&=&\frac{i k}{a} \textmd{ }\textmd{ }\textmd{ } \textmd{with}
\textmd{ } \textmd{ }\textmd{ }   k=\sqrt{E^2-m_0^2}\\\label{9}
\nu&=&\frac{1}{2}+\frac{1}{2}\sqrt{1-\frac{4}{a^2q^2}[V_0^2-(m_1+S_0)^2]}\\\label{10}
\gamma&=&\frac{1}{a}\sqrt{\left(m_0-\frac{m_1+S_0}{q}\right)^2-\left(E+\frac{V_0}{q}\right)^2}\label{11}.
\end{eqnarray}
It is worth noting that $|E|>m$ ensures that $k$ is real and $V_0$
is real and positive for the scattering states [9, 10]. One can
write the solution of the Eq.~(\ref{8}) in terms of hypergeometric
functions [33]
\begin{eqnarray}
H(y)&=&A_1\textmd{
}F(\mu+\nu-\gamma,\mu+\nu+\gamma,1+2\mu,y)\nonumber\\&+&A_2
y^{-2\mu}\textmd{ }F(-\mu+\nu-\gamma,-\mu+\nu+\gamma,1-2\mu,y)\,.
\end{eqnarray}
Thus, the left solution is obtained as follows
\begin{eqnarray}
\Psi_L(y)&=&A_1 y^{\mu}(1-y)^{\nu}\textmd{ }F(\mu+\nu-\gamma,\mu+\nu+\gamma,1+2\mu,y)\nonumber\\
&+&A_2 y^{-\mu}(1-y)^{\nu} \textmd{
}F(-\mu+\nu-\gamma,-\mu+\nu+\gamma,1-2\mu,y).\label{12}
\end{eqnarray}
Let us investigate the scattering solution for $x>0$. In that
case, Eq.~(\ref{2}) becomes
\begin{equation}
\frac{d^2\Psi_R(x)}{d
x^2}+\left\{\left[E-\frac{V_0}{e^{bx}-\tilde{q}}\right]^2-\left[m_0+\frac{m_1+S_0}{e^{bx}-\tilde{q}}\right]^2\right\}\Psi_R(x)=0\label{13}.
\end{equation}
Changing the variable $e^{bx}=\tilde{q}/z$ and setting
$\Psi_{R}(z)=z^{\delta}(1-z)^{-\alpha}G(z)$, Eq.~(\ref{13}) yields
\begin{equation}
z(1-z)\frac{d^2G(z)}{dz^2}+[2\delta+1-(2\delta-2\alpha+1)z]\frac{dG(z)}{dz}-(\delta-\alpha-\beta)(\delta-\alpha+\beta)G(z)=0\label{14}
\end{equation}
where
\begin{eqnarray}
\delta&=&\frac{i k}{b}\\\label{15}
\alpha&=&-\frac{1}{2}+\frac{1}{2}\sqrt{1-\frac{4}{b^2\tilde{q}^2}[V_0^2-(m_1+S_0)^2]}\\\label{16}
\beta&=&\frac{1}{b}\sqrt{\left(m_0-\frac{m_1+S_0}{\tilde{q}}\right)^2-\left(E+\frac{V_0}{\tilde{q}}\right)^2}\label{17}.
\end{eqnarray}
Eq.~(\ref{14}) is the hypergeometric equation and its general
solution is given as [33]
\begin{eqnarray}
G(z)&=&A_3\textmd{
}F(\delta-\alpha-\beta,\delta-\alpha+\beta,1+2\delta,z)\nonumber\\&+&A_4
z^{-2\delta}\textmd{
}F(-\delta-\alpha-\beta,-\delta-\alpha+\beta,1-2\delta,z).
\end{eqnarray}
Finally, the right solution can be written as
\begin{eqnarray}
\Psi_R(z)&=&A_3 z^{\delta}(1-z)^{-\alpha}\textmd{ }F(\delta-\alpha-\beta,\delta-\alpha+\beta,1+2\delta,z)\nonumber\\
&+&A_4 z^{-\delta}(1-z)^{-\alpha} \textmd{
}F(-\delta-\alpha-\beta,-\delta-\alpha+\beta,1-2\delta,z).\label{18}
\end{eqnarray}

\section{Transmission and reflection coefficients}

In order to obtain the transmission and reflection coefficients,
we have to investigate the asymptotic behavior of the left and
right solutions. As $x \rightarrow - \infty$, then $y \rightarrow
0$, $(1-y)^{\nu} \rightarrow 1$ and $y^{\pm \mu} \rightarrow
q^{\pm \mu}e^{\pm a\mu x}$ which leads to
\begin{equation}
\Psi_L(x\rightarrow - \infty)\sim A_1 q^{\mu}e^{i k x} + A_2
q^{-\mu}e^{- i k x} \label{19}
\end{equation}
where following property of the hypergeometric functions is used:
$_2F_1(a,b;c;0)=1$ [33]. Asymptotic behavior of the left solution
can be written in terms of incident $\Psi_{inc}$ and reflected
$\Psi_{ref}$ waves in the limit $x \rightarrow - \infty$. Then, it
is seen from Eq.~({\ref{19}}) that $\Psi_{inc}$ and $\Psi_{ref}$
behave like a plane wave travelling to the right and left,
respectively.

On the other hand, as $x \rightarrow \infty$, then $z \rightarrow
0$, $(1-z)^{-\alpha} \rightarrow 1$ and $z^{\pm \delta}
\rightarrow \tilde{q}^{\pm \delta}e^{\mp b\delta x}$ which leads
to
\begin{equation}
\Psi_R(x\rightarrow \infty)\sim A_3 q^{\delta}e^{- b\delta x} +
A_4 q^{-\delta}e^{b\delta x}. \label{20}
\end{equation}
To obtain the transmitted wave traveling from left to right, we
set $A_3=0$. So, we get
\begin{equation}
\Psi_R(x\rightarrow \infty)\sim  A_4 q^{-\delta}e^{i k x}.
\label{21}
\end{equation}
Thus, from now on, the following right solution is used:
\begin{equation}
\Psi_R(z)=A_4 z^{-\delta}(1-z)^{-\alpha} \textmd{
}F(-\delta-\alpha-\beta,-\delta-\alpha+\beta,1-2\delta,z).\label{21a}
\end{equation}
Then, the transmission and reflection coefficients can be
expressed as [11]
\begin{equation}
T=\left|\frac{\Psi_{trans}}{\Psi_{inc}}\right|^2=\left|\frac{A_4}{A_1}\right|^2
\label{22}
\end{equation}
\begin{equation}
R=\left|\frac{\Psi_{ref}}{\Psi_{inc}}\right|^2=\left|\frac{A_2}{A_1}\right|^2\label{23}.
\end{equation}
Let's use the following continuity conditions on the wave
functions and their first derivatives at $x=0$ to obtain explicit
expressions for $R$ and $T$
\begin{eqnarray}
\Psi_R(x=0)&=&\Psi_L(x=0)\label{24}\\
\frac{d}{dx}\Psi_R(x=0)&=&\frac{d}{dx}\Psi_L(x=0).\label{25}
\end{eqnarray}
Using Eqs.~({\ref{24}}) and~({\ref{25}}) with Eqs.~({\ref{12}})
and~({\ref{18}}), one can obtain
\begin{eqnarray}
A_4 C_1 F_1 &=&A_1 C_2 F_2 + A_2 C_3 F_3\label{26},\\
A_4 C_1 b (C_4 F_4-C_5 F_1)&=& A_1 C_2 a (C_6 F_2+C_7 F_5)-A_2 C_3
a (C_8 F_3-C_9 F_6)\label{27}
\end{eqnarray}
where we have used the following property: $\frac{d}{dx}\textmd{
}F(a,b;c;x)=\frac{a b}{c}\textmd{ }F(a+1,b+1;c+1;x)$ [33]. Here,
some new coefficients and definitions are used to shorten the
Eqs.~({\ref{26}}) and ({\ref{27}}) and they are listed in Table 1.
Consequently, transmission and reflection coefficients are,
respectively, calculated as follows
\begin{equation}
T=\left|\frac{A_4}{A_1}\right|^2=\left|\frac{a C_2[F_2(C_8 F_3-C_9
F_6)+ F_3(C_6 F_2+C_7 F_5)]}{C_1[aF_1(C_8 F_3-C_9 F_6)+bF_3(C_4
F_4-C_5 F_1)]}\right|^2\label{28}
\end{equation}
\begin{equation}
R=\left|\frac{A_2}{A_1}\right|^2=\left|\frac{C_2[aF_1(C_6 F_2+C_7
F_5)-bF_2(C_4 F_4-C_5 F_1)]}{C_3[aF_1(C_8 F_3-C_9 F_6)+ bF_3(C_4
F_4-C_5 F_1)]}\right|^2.\label{29}
\end{equation}
It is significant to discuss the condition for existence of the
transmission resonances in the scattering states. The transmission
resonance condition can be derived by setting $R=0$ (or $T=1$)
which means that there is no reflected wave. Then, considering
Eq.~({\ref{29}}), the condition for the transmission resonances is
obtained as
\begin{equation}
a F_1 \left[C_6 F_2 +C_7 F_5\right]-b F_2 \left[C_4 F_4 -C_5
F_1\right]=0.\label{29a}
\end{equation}

\section{Bound states}
In this section, we investigate the bound state solution of the
effective mass Klein-Gordon equation for the asymmetric
Hulth{\'e}n potential. The scalar~({\ref{3}}) and
vector~({\ref{4}}) asymmetric Hulth{\'e}n potentials give a bound
state $|E|<m$ as they become attractive ($V_0\rightarrow -V_0$ and
$S_0\rightarrow -S_0$) [34]. Then, in the bound state case,
Eq.~({\ref{2}}) yields the following second-order differential
equations for $x<0$ and $x>0$, respectively
\begin{equation}
\frac{d^2\Phi_L(x)}{d
x^2}+\left\{\left[E+\frac{V_0}{e^{-ax}-q}\right]^2-\left[m_0+\frac{m_1-S_0}{e^{-ax}-q}\right]^2\right\}\Phi_L(x)=0\label{32}
\end{equation}
\begin{equation}
\frac{d^2\Phi_R(x)}{d
x^2}+\left\{\left[E+\frac{V_0}{e^{bx}-\tilde{q}}\right]^2-\left[m_0+\frac{m_1-S_0}{e^{bx}-\tilde{q}}\right]^2\right\}\Phi_R(x)=0\label{38}.
\end{equation}
Following the same procedure as above, we obtain the solutions of
the Klein-Gordon equation for both $x<0$ and $x>0$ in terms of
hypergeometric functions as follow
\begin{eqnarray}
\Phi_L(y)&=&B_1 y^{\tilde{\mu}}(1-y)^{\tilde{\nu}}
\textmd{ }F(\tilde{\mu}+\tilde{\nu}-\tilde{\gamma},\tilde{\mu}+\tilde{\nu}+\tilde{\gamma},1+2\tilde{\mu},y)\nonumber\\
&+&B_2 y^{-\tilde{\mu}}(1-y)^{\tilde{\nu}} \textmd{
}F(-\tilde{\mu}+\tilde{\nu}-\tilde{\gamma},-\tilde{\mu}
+\tilde{\nu}+\tilde{\gamma},1-2\tilde{\mu},y)\label{37}
\end{eqnarray}
\begin{eqnarray}
\Phi_R(z)&=&B_3 z^{\tilde{\delta}}(1-z)^{-\tilde{\alpha}}
\textmd{ }F(\tilde{\delta}-\tilde{\alpha}-\tilde{\beta},\tilde{\delta}-\tilde{\alpha}+\tilde{\beta},1+2\tilde{\delta},z)\nonumber\\
&+&B_4 z^{-\tilde{\delta}}(1-z)^{-\tilde{\alpha}} \textmd{
}F(-\tilde{\delta}-\tilde{\alpha}-\tilde{\beta},-\tilde{\delta}-\tilde{\alpha}+\tilde{\beta},1-2\tilde{\delta},z)\label{43}
\end{eqnarray}
where
\begin{eqnarray}
\tilde{\mu}=\frac{1}{a}\sqrt{m_0^2-E^2}, \textmd{ }
\textmd{ }\textmd{ }\tilde{\delta}=\frac{1}{b}\sqrt{m_0^2-E^2}\textmd{ }\textmd{ }\textmd{ }\textmd{ }\textmd{ }\textmd{ }
\textmd{ }\textmd{ }\textmd{ }\textmd{ }\textmd{ }\textmd{ }\textmd{ }\textmd{ }\textmd{ }\textmd{ }\textmd{ }\textmd{ }\textmd{ }
\textmd{ }\textmd{ }\textmd{ }\textmd{ }\textmd{ }\textmd{ }\textmd{ }\textmd{ }\textmd{ }\textmd{ }\textmd{ }\textmd{ }\textmd{ }\textmd{ }\textmd{ }\label{34}\\
\tilde{\nu}=\frac{1}{2}\left(1+\sqrt{1-\frac{4}{a^2q^2}[V_0^2-(m_1-S_0)^2]}\right), \textmd{ }
\tilde{\alpha}=\frac{1}{2}\left(-1+\sqrt{1-\frac{4}{b^2\tilde{q}^2}[V_0^2-(m_1-S_0)^2]}\right)\label{35}\\
\tilde{\gamma}=\frac{1}{a}\sqrt{\left(m_0-\frac{m_1-S_0}{q}\right)^2-\left(E-\frac{V_0}{q}\right)^2},
\textmd{ }
\tilde{\beta}=\frac{1}{b}\sqrt{\left(m_0-\frac{m_1-S_0}{\tilde{q}}\right)^2-\left(E-\frac{V_0}{\tilde{q}}\right)^2}\label{36}.
\end{eqnarray}
In order to obtain the bound state solutions requiring the
vanishing of the wave functions~({\ref{37}}) and ~({\ref{43}}) at
$\pm \infty$, the following regular solutions are chosen for
$\Phi_L(y)$ and $\Phi_R(z)$, respectively
\begin{eqnarray}
\Phi_L(y)&=&B_1 y^{\tilde{\mu}}(1-y)^{\tilde{\nu}} \textmd{
}F(\tilde{\mu}+\tilde{\nu}-\tilde{\gamma},\tilde{\mu}+\tilde{\nu}+\tilde{\gamma},1+2\tilde{\mu},y)\label{44}
\end{eqnarray}
\begin{eqnarray}
\Phi_R(z)&=&B_3 z^{\tilde{\delta}}(1-z)^{-\tilde{\alpha}} \textmd{
}F(\tilde{\delta}-\tilde{\alpha}-\tilde{\beta},\tilde{\delta}-\tilde{\alpha}+\tilde{\beta},1+2\tilde{\delta},z)\label{45}.
\end{eqnarray}
Imposing the continuity conditions on  $\Phi_L(y)$ and $\Phi_R(z)$
at $x=0$, we obtain
\begin{eqnarray}
B_1 \tilde{C_1} \tilde{F_1} - B_3 \tilde{C_2} \tilde{F_2}&=&0\label{46}\\
B_1 a \tilde{C_3} (\tilde{C_4} \tilde{F_1}+\tilde{C_5}
\tilde{F_3})- B_3 b \tilde{C_6} (\tilde{C_7}
\tilde{F_2}+\tilde{C_8} \tilde{F_4})&=&0.\label{47}
\end{eqnarray}
Here, we use some abbreviations given in Table 2. From the last
two equations, one can calculate the energy eigenvalue equation as
follows
\begin{equation}
a\tilde{C_2}\tilde{C_3}\tilde{F_2}(\tilde{C_4}\tilde{F_1}+\tilde{C_5}
\tilde{F_3})-b\tilde{C_1}\tilde{C_6}\tilde{F_1}(\tilde{C_7}\tilde{F_2}+\tilde{C_8}\tilde{F_4})=0.\label{48}
\end{equation}
The energy eigenvalues of the bound states can be obtained in
terms of potential strength $V_0$ by solving the Eq.~({\ref{48}})
numerically. The dependence of the energy eigenvalues on $V_0$ is
given in Fig. 2. From Fig. 2, one can observe that bound state
energy of the Klein-Gordon particle decreases with increasing
potential strength $V_0$. Finally, this energy takes the value of
negative mass of the particle which means that the bound state
joins the negative-energy continuum [35, 14]. This result agrees
with previous ones [11, 14]. On the other hand, the potential is
called critical when a discrete energy level crosses the value
$E=-m$ and joins the negative energy continuum.

\section{Discussions}
\subsection{Low-momentum limit}
In the low-momentum limit $(E\rightarrow -m)$ which leads to
$\mu=\delta=0$, transmission resonance condition~({\ref{29a}})
becomes
\begin{eqnarray}
 a\textmd{ }F(-\alpha-\beta,-\alpha+\beta,1,\tilde{q}) \times \nonumber\\
 \left[q(\nu^2-\gamma^2)\textmd{ }F(\nu-\gamma+1,\nu+\gamma+1,2,q)-\frac{q\nu}{1-q}\textmd{ }F(\nu-\gamma,\nu+\gamma,1,q)\right] -\nonumber\\
b\textmd{ }F(\nu-\gamma,\nu+\gamma,1,q) \times \nonumber\\
\left[\tilde{q}(\beta^2-\alpha^2) \textmd{
}F(1-\alpha-\beta,1-\alpha+\beta,2,\tilde{q})-\frac{\tilde{q}\alpha}{1-\tilde{q}}
\textmd{
}F(-\alpha-\beta,-\alpha+\beta,1,\tilde{q})\right]=0.\label{49}
\end{eqnarray}
On the other hand, bound state energy equation~({\ref{48}}) turns
into the following form in the low-momentum limit
 ($\tilde{\mu}=\tilde{\delta}=0$):
\begin{eqnarray}
a\textmd{
}F(-\tilde{\alpha}-\tilde{\beta},-\tilde{\alpha}+\tilde{\beta},1,\tilde{q})
\times \nonumber\\
\left[q(\tilde{\nu}^2-\tilde{\gamma}^2)\textmd{
}F(\tilde{\nu}-\tilde{\gamma}+1,
\tilde{\nu}+\tilde{\gamma}+1,2,q)-\frac{q\tilde{\nu}}{1-q}\textmd{ }F(\tilde{\nu}-\tilde{\gamma},\tilde{\nu}+\tilde{\gamma},1,q)\right] -\nonumber\\
b\textmd{
}F(\tilde{\nu}-\tilde{\gamma},\tilde{\nu}+\tilde{\gamma},1,q)
\times \nonumber\\
\left[\tilde{q}(\tilde{\beta}^2-\tilde{\alpha}^2) \textmd{
}F(1-\tilde{\alpha}-\tilde{\beta},1-\tilde{\alpha}+\tilde{\beta},2,\tilde{q})-\frac{\tilde{q}\tilde{\alpha}}{1-\tilde{q}}
\textmd{
}F(-\tilde{\alpha}-\tilde{\beta},-\tilde{\alpha}+\tilde{\beta},1,\tilde{q})\right]=0.\label{50}
\end{eqnarray}
Comparing Eq.~({\ref{49}}) with Eq.~({\ref{50}}) and considering
Eqs.~({\ref{9}}), ({\ref{11}}), ({\ref{15}}) and ({\ref{17}}), it
is not difficult to see that transmission resonance condition is
reduced to the bound-energy condition after the transformations
$V_0 \rightarrow -V_0$ and $S_0 \rightarrow -S_0$ in the
low-momentum limit which means that the
asymmetric Hulth{\'e}n potential supports a zero-momentum (half-bound) state in the presence of the effective mass and also constant mass as well.\\

Eq.~({\ref{50}}) can be used to calculate the value of the
critical potential $(E=-m)$. This value is found to be
$V_c=1.89014$ for $a = 1$, $b = 0.8$, $q = 0.5$, $\tilde{q} =
0.4$, $m = 1$, $ m_1= 0$, $S = 0$ and $V_c=2.0512 $ for $a = 1$,
$b = 0.8$, $q = 0.5$, $\tilde{q} = 0.4$, $m = 1$, $ m_1= 0.2$, $S
= 0$. From this point of view, we can say that  the value of the
critical potential increases in the presence of the effective mass
compared to the constant mass case.

\subsection{Unitary condition}
Fig. 3 shows the existence of the unitary condition for the
asymmetric Hulth{\'e}n potential within the effective mass
formalism. From Fig. 3, we can see that unitary condition
($T+R=1$) is valid for both position dependent (right plot) and
constant (left plot) mass cases.

\subsection{Transmission resonances}
 Figs. 4-9 display the transmission coefficients for the asymmetric Hulth{\'e}n potential. All of the figures show that
transmission resonances for the asymmetric Hulth{\'e}n potential
exist for both of the effective mass and constant mass cases. Fig.
4 presents the behavior of the transmission coefficient versus the
scalar particles energy. In Fig. 4, solid line represents the
constant mass case for the usual Hulth{\'e}n potential displayed
in Fig. 3 in Ref. [18]. We can readily see from Fig. 4 that the
width of the transmission resonances is sensitive to the effective
mass parameter $m_1$. The peaks of the resonances become narrower
when $m_1$ increases. Besides, in the case of the effective mass,
the first resonance peak ($T=1$) appears at lower energy compared
to the constant mass case.

Fig. 5 displays the transmission coefficients as a function of
potential strength $V_0$. In Ref. [13], it is shown that
transmission resonances for the Klein-Gordon particles in the
presence of the Woods-Saxon potential vanishes for $E-m<V_0<E+m$
and they appear for $V_0>E+m$. However, from Fig. 5, we can see
that transmission resonances appear all range of the asymmetric
Hulth{\'e}n potential which means that there is no $V_0$ values
making the asymmetric Hulth{\'e}n potential entirely impenetrable.
The reason is that there is no way to reduce the asymmetric
Hulth{\'e}n potential to square well. Thus, asymmetric Hulth{\'e}n
potential for the Klein-Gordon particle is completely penetrable
as in the vector particle case [32]. The result given in Fig. 5
with solid line ($m_1=0$ constant mass) also agrees with the one
presented in Fig. 4 in Ref. [18]. Effects of the
position-dependent mass on the transmission coefficients are
represented with the other three lines. From these three lines, it
is easy to conclude that transmission resonances also appear in
all range of the asymmetric Hulth{\'e}n potential in the presence
of the effective mass and resonance peaks become narrower and
shorter with increasing the $m_1$.

\subsection{Effect of the potential parameters on the transmission resonances}
From the Figs. 6-8, it is clear that intensities of the
transmission resonances as well as the width of the resonance
peaks depend on the shape of the external potential. From the Fig.
6, one can observe  that the dependence of the transmission
coefficient on the energy of the Klein-Gordon particle is the same
for both $a>b$ and $b>a$. The reason can be found by considering
the Fig. 1. Based on the left plot of the Fig. 1, one can readily
notice that the height and the width of the potential barrier
remains the same whether $a>b$ or $b>a$. However, comparing the
Fig. 6 with the Fig. 4, we can conclude that the intensity of the
resonance peaks increases with decreasing $a$ (or $b$) for
$q=\tilde{q}$. On the other hand, the first transmission resonance
peak appears at smaller values of the Klein-Gordon particles
energy in the presence of the position-dependent mass.

Relationship between potential parameters $q$ and $\tilde{q}$ that
define the shape of the potential~({\ref{1}}) and transmission
coefficient is given in the Fig. 7.  From the Fig. 7, we can see
that existence of the transmission resonances depends on $q$ and
$\tilde{q}$. However, it should be noticed that the form of the
transmission resonances remain the same in both $q > \tilde{q}$
and $q < \tilde{q}$. This can be explained by considering the
change in the hight and the width of asymmetric Hulth{\'e}n
potential as shown in middle plot of the Fig. 1. From this plot,
we can conclude that magnitudes of the hight and the width of the
asymmetric Hulth{\'e}n potential remains the same for $q >
\tilde{q}$ and $q < \tilde{q}$. However, by considering the Figs.
7 and 8, it is seen that the number of the transmission resonances
increases with increasing the $q$ and $\tilde{q}$.

We also investigate the energy dependence of the transmission
resonances for both $a>b>q>\tilde{q}$ and $a<b<q<\tilde{q}$. The
results are given in the Fig. 8. From the left plot of the Fig. 8,
it is concluded that the intensities and widths of the resonance
peaks decrease as the potential parameters $a$ and $b$ are bigger
than the $q$ and $\tilde{q}$. The reason can be found by
considering the right plot of the Fig. 1. This plot gives that the
hight and the width of the asymmetric Hulth{\'e}n
potential~({\ref{1}}) decrease when $a > b > q > \tilde{q}$. On
the other hand, the Fig. 8 shows that if the $a$ and $b$ are
smaller than the $q$ and $\tilde{q}$, then the number of
transmission resonance peaks increases. Based on the Figs. 5-8, we
can conclude that existence of the transmission resonances as well
as the intensity and width of the resonance peaks depend on shape
of the asymmetric Hulth{\'e}n potential. Besides, we can also see
from the Figs. 5-8 that the Klein-Gordon equation within the
position-dependent mass formalism exhibits transmission resonance
in the presence of the asymmetric Hulth{\'e}n potential.

\subsection{Effect of the unequal scalar and vector potentials on the transmission resonances}
The Fig. 9 displays the effect of the unequal scalar and vector
potentials on the transmission resonances. From the Fig. 9, one
can observe that transmission resonance peaks disappear when
$S_0\neq0$ and it does not matter whether $V_0=S_0$ or $V_0<S_0$.

\section{Conclusions}
In the present study, exact solution of the one-dimensional
effective mass Klein-Gordon equation for the asymmetric
Hulth{\'e}n potential has been found in terms of hypergeometric
functions. Considering the asymptotic behavior of the
hypergeometric functions and using the continuity of the wave
functions, we have obtained the scattering and bound states of the
Klein-Gordon particle. Then, the condition for the existence of
the transmission resonances $(T=1, R=0)$ is derived for the
position-dependent mass and constant mass, as a special case. From
the Figs. 4-8, it has been observed that the intensity and width
of transmission resonance peaks depend on the shape and the
strength $V_0$ of the external potential. Based on the Figs. 3-8,
we have concluded that asymmetric and symmetric Hulth{\'e}n
potentials are entirely penetrable for all values of potential
strength $V_0$ since asymmetric and symmetric Hulth{\'e}n
potentials cannot be reduced to square well. In the low-momentum
limit, it has been shown that asymmetric Hulth{\'e}n potential
supports zero-momentum (half-bound) states. On the other hand, our
results show that the Klein-Gordon equation exhibits the
transmission resonances in the presence of the position-dependent
mass and form of the resonances depends on the effective mass
parameter $m_1$  as well as the shape and the strength of the
external potential. Furthermore, asymmetric Hulth{\'e}n potential
has a general form and reduces to the well-known potential such as
usual Hulth{\'e}n and Cusp potentials. Thus, our results contain
the scattering and bound state solutions of the scalar particles
for the Cusp and usual Hulth{\'e}n potentials. Finally, it should
be mentioned that these results can be useful for understanding
the behavior of elementary particles and nuclei in view of
effective mass.

\section{Acknowledgments}
This research was partially supported by the Scientific and
Technical Research Council of Turkey.

\newpage

\begin{table}
\caption{Table for coefficients and definitions used to calculate
the T and R. }
\begin{tabular}{cccc}
\hline
$C_1$&$\tilde{q}^{-\delta}(1-\tilde{q})^{-\alpha}$&$C_6$&$\mu-\frac{\nu q}{1-q}$\\
$C_2 $&$ q^{\mu} (1-q)^{\nu}$&$C_7$&$q\frac{(\mu+\nu-\gamma)(\mu+\nu+\gamma)}{1+2\mu}$ \\
$C_3 $&$ q^{-\mu} (1-q)^{\nu}$&$C_8$&$\mu+\frac{\nu q}{1-q}$\\
$C_4$&$\tilde{q}\frac{(\delta+\alpha+\beta)(\beta-\delta-\alpha)}{1-2\delta}$&$C_9$&$q\frac{(\nu-\mu-\gamma)(\nu-\mu+\gamma)}{1-2\mu}$\\
$C_5$&$\frac{\alpha \tilde{q}}{1-\tilde{q}}-\delta$\\
$F_1$&$\textmd{ }F(-\delta-\alpha-\beta,-\delta-\alpha+\beta;1-2\delta;\tilde{q})$&$F_4$&$\textmd{ }F(1-\delta-\alpha-\beta,1-\delta-\alpha+\beta;2-2\delta;\tilde{q})$\\
$F_2$&$\textmd{ }F(\mu+\nu-\gamma,\mu+\nu+\gamma;1+2\mu;q)$&$F_5$&$\textmd{ }F(\mu+\nu-\gamma+1,\mu+\nu+\gamma+1;2+2\mu;q)$\\
$F_3$&$\textmd{ }F(-\mu+\nu-\gamma,-\mu+\nu+\gamma;1-2\mu;q)$&$F_6$&$\textmd{ }F(1-\mu+\nu-\gamma,1-\mu+\nu+\gamma;2-2\mu;q)$\\
\hline
\end{tabular}
\end{table}

\begin{table}
\caption{Table for coefficients and definitions used to calculate
the bound-state energy. }
\begin{tabular}{cccc}
\hline
$\tilde{C_1}$&$q^{\tilde{\mu}} (1-q)^{\tilde{\nu}}$& $\tilde{C_5}$& $\frac{(\tilde{\mu}+\tilde{\nu}-\tilde{\gamma})(\tilde{\mu}+\tilde{\nu}+\tilde{\gamma})}{1+2\tilde{\mu}}$\\
$\tilde{C_2}$&$\tilde{q}^{\tilde{\delta}}(1-\tilde{q})^{-\tilde{\alpha}}$& $\tilde{C_6}$&$-\tilde{q}^{\tilde{\delta}+1}(1-\tilde{q})^{-\tilde{\alpha}}$\\ $\tilde{C_3}$&$q^{\tilde{\mu}+1} (1-q)^{\tilde{\nu}}$&$\tilde{C_7}$&$\frac{\tilde{\alpha}}{1-\tilde{q}}+\frac{\tilde{\delta}}{\tilde{q}}$\\
$\tilde{C_4}$&$\frac{\tilde{\mu}}{q}-\frac{\tilde{\nu}}{1-q}$ &$\tilde{C_8}$&$\frac{(\tilde{\delta}-\tilde{\alpha}-\tilde{\beta})(\tilde{\beta}+\tilde{\delta}-\tilde{\alpha})}{1+2\tilde{\delta}}$\\
$\tilde{F_1}$&$\textmd{ }F(\tilde{\mu}+\tilde{\nu}-\tilde{\gamma},\tilde{\mu}+\tilde{\nu}+\tilde{\gamma};1+2\tilde{\mu};q)$ & $\tilde{F_3}$&$\textmd{ }F(\tilde{\mu}+\tilde{\nu}-\tilde{\gamma}+1,\tilde{\mu}+\tilde{\nu}+\tilde{\gamma}+1;2+2\tilde{\mu};q)$\\
$\tilde{F_2}$&$\textmd{
}F(\tilde{\delta}-\tilde{\alpha}-\tilde{\beta},\tilde{\delta}-\tilde{\alpha}+\tilde{\beta};1+2\tilde{\delta};\tilde{q})$
& $\tilde{F_4}$&$\textmd{ }F(1+\tilde{\delta}-\tilde{\alpha}-\tilde{\beta},1+\tilde{\delta}-\tilde{\alpha}+\tilde{\beta};2+2\tilde{\delta};\tilde{q})$\\
\hline
\end{tabular}
\end{table}
\newpage
$\left.\textmd{ }\right.$
\newpage

\begin{figure}
\caption{The form of the asymmetric Hulth{\'e}n potential with
$V_0=1$. }
    \begin{center}
      \resizebox{180mm}{!}{\includegraphics[angle=0]{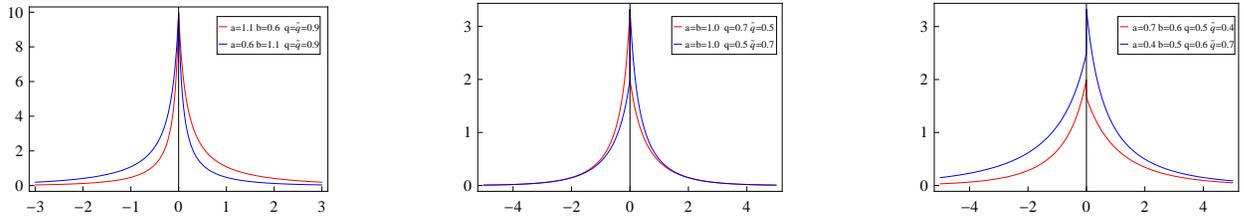}}
            \end{center}
      \end{figure}

\begin{figure}
\caption{Energy of the lowest bound-state vs potential strength
 in the presence of the position-independent(solid line)/dependent (dashed line) mass cases. }
    \begin{center}
      \resizebox{100mm}{!}{\includegraphics[angle=0]{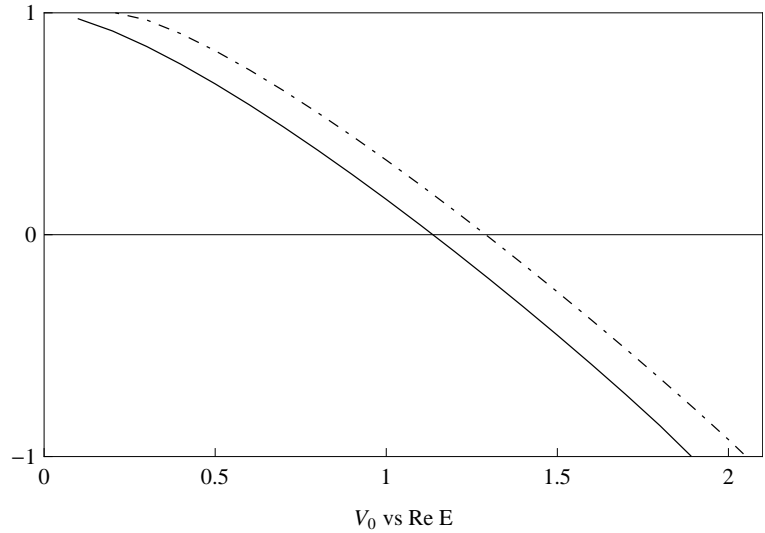}}
    \end{center}
      \end{figure}

\begin{figure}
\caption{T (solid line) and R (dashed line) coefficients for
$a=0.8$, $q=0.5$, $b=0.9$, $\tilde{q}=0.6$, $m_0=1$, $S_0=0$ and
$V_0=4$ where $m_1=0.0$ and $m_1=0.5$ for the left and right
plots, respectively. }
    \begin{center}
      \resizebox{170mm}{!}{\includegraphics[angle=0]{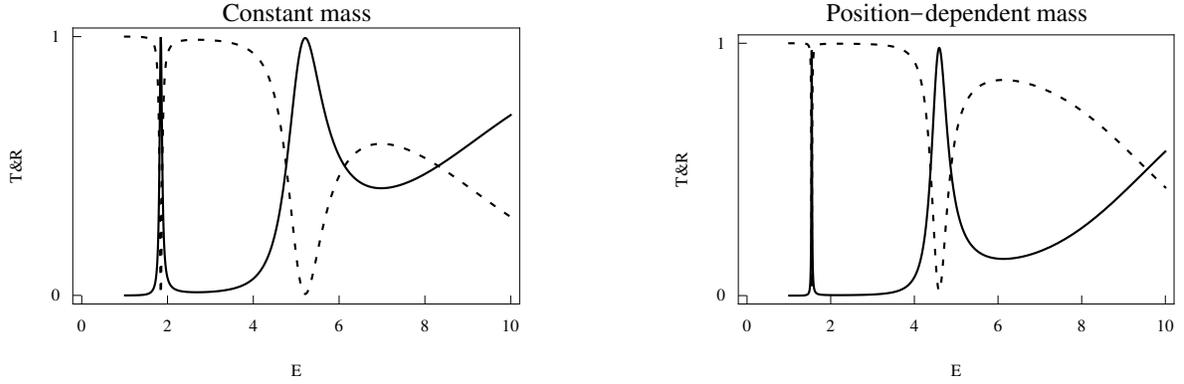}}
    \end{center}
      \end{figure}

\begin{figure}
\caption{Transmission coefficients for constant mass (solid line)
and position-dependent mass (dashed line) with $q=0.9$,
$\tilde{q}=0.9$, $m_0=1$, $m_1=0.5$, $S_0=0$ and $V_0=4$. We also
take $a=1.1$ and $b=0.6$ for the left plot and $a=0.6$ and $b=1.1$
for the right plot. }
  \begin{center}
      \resizebox{120mm}{!}{\includegraphics[angle=0]{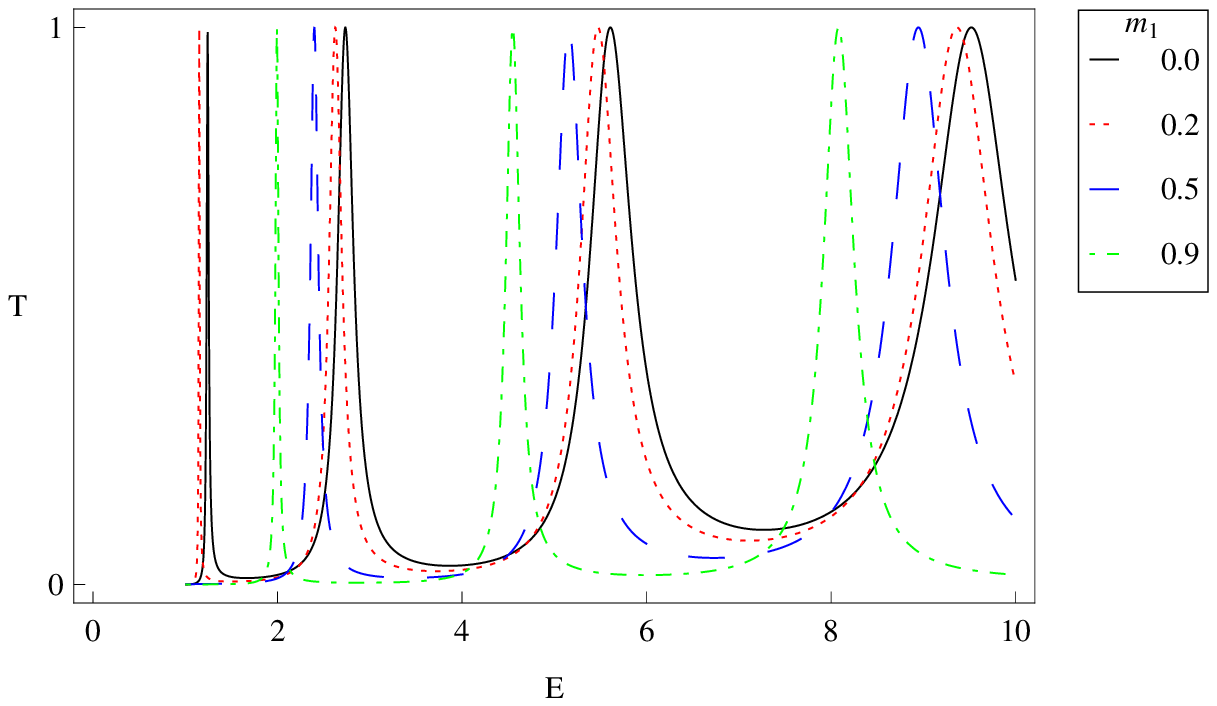}}
        \end{center}
  \end{figure}

\begin{figure}
\caption{Transmission coefficient vs energy for $a = 1$, $b = 1$,
$q = 0.9$, $\tilde{q} = 0.9$, $m_0 = 1$, $S_0 = 0$, $V_0 = 4$. }
   \begin{center}
      \resizebox{130mm}{!}{\includegraphics[angle=0]{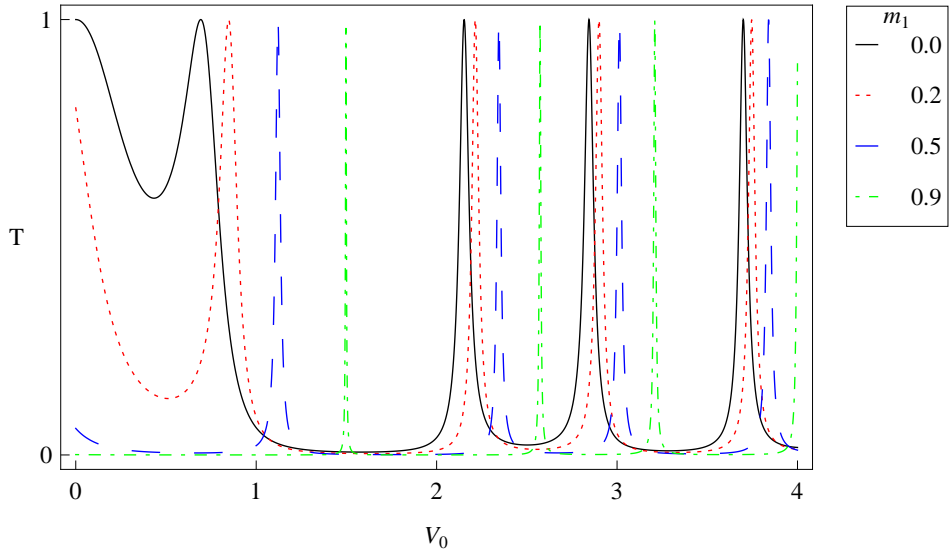}}
    \end{center}
      \end{figure}

\begin{figure}
\caption{Transmission coefficient vs potential strength for $a =
1$, $b = 1$, $q = 0.9$, $\tilde{q} = 0.9$, $m_0 = 1$, $S_0 = 0$,
$E = 2m_0$. }
    \begin{center}
      \resizebox{170mm}{!}{\includegraphics[angle=0]{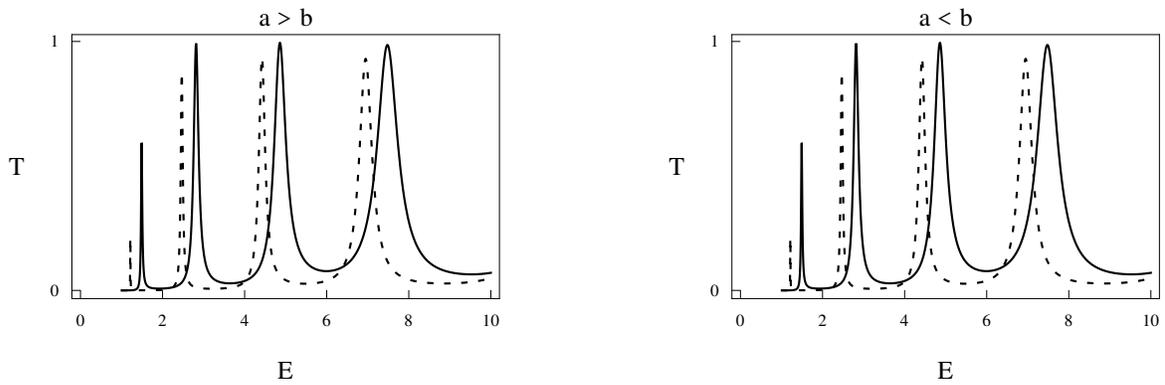}}
    \end{center}
      \end{figure}

\begin{figure}
\caption{Transmission coefficients for constant mass (solid line)
and position-dependent mass (dashed line) with $a=1$, $b=1$,
$m_0=1$, $m_1=0.5$, $S_0=0$ and $V_0=4$. We also take $q=0.6$ and
$\tilde{q}=0.5$ for the left plot and $q=0.5$ and $\tilde{q}=0.6$
for the right plot. }
    \begin{center}
      \resizebox{170mm}{!}{\includegraphics[angle=0]{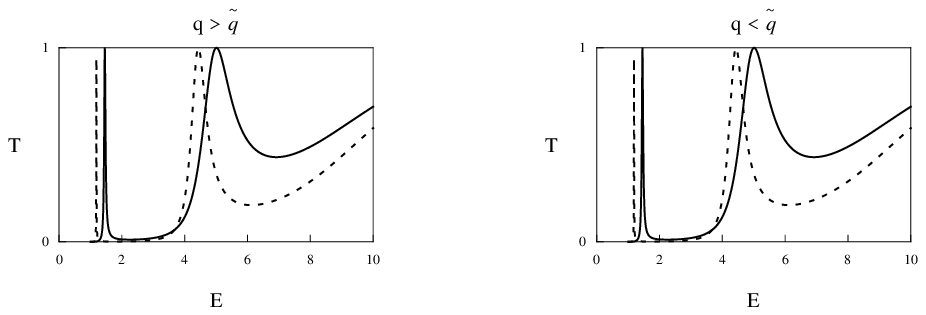}}
    \end{center}
      \end{figure}

\begin{figure}
\caption{Transmission coefficients for constant mass (solid line)
and position-dependent mass (dashed line) with $a=0.7$, $b=0.6$,
$q=0.5$, $\tilde{q}=0.4$, $m_0=1$, $m_1=0.5$, $S_0=0$ and $V_0=4$
for the left plot and $a=0.4$, $b=0.5$, $q=0.6$, $\tilde{q}=0.7$,
$m_0=1$, $m_1=0.5$, $S_0=0$ and $V_0=4$ for the right plot. }
    \begin{center}
      \resizebox{170mm}{!}{\includegraphics[angle=0]{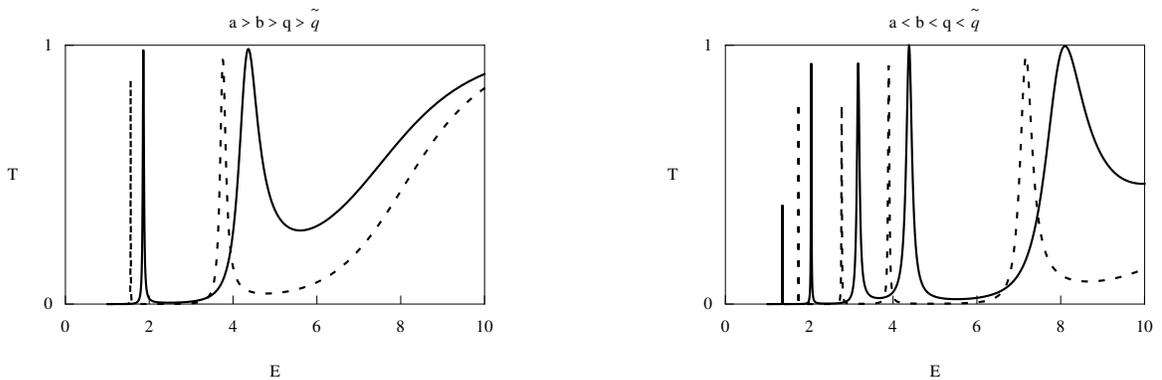}}
    \end{center}
      \end{figure}

\begin{figure}
\caption{Transmission coefficients in the constant mass (solid
line) case and position-dependent mass (dashed line) case with
$a=b=q=\tilde{q}=0.5$, $m_0=1$, $m_1=0.5$ for $S_0=V_0=0.5$ (the
left plot) and for $S_0=0.4$, $V_0=0.6$ (the right plot). }
    \begin{center}
      \resizebox{170mm}{!}{\includegraphics[angle=0]{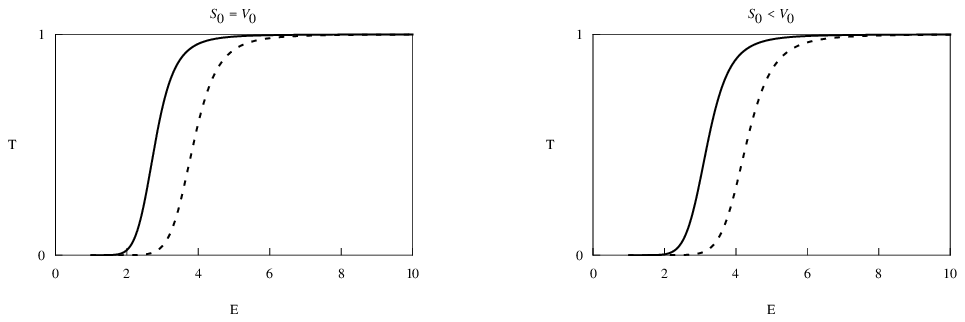}}
    \end{center}
      \end{figure}


\newpage



\begin{thebibliography}{99}

\bibitem{1} R. G. Newton, Scattering Theory of Waves and Particles, Springer-Verlag, Berlin, 1982.
\bibitem{2} L. D. Faddeev, Properties of the S-matrix of the one-dimensional Schrodinger equation, Trudy Mat. Inst. Stekl. 73 (1964) 314-336.
\bibitem{3} Y. Jiang, S. H. Dong, A. Antillon and M. Lozada-Cassou, Low momentum scattering of
the Dirac particlewith an asymmetric cusp potential, Eur. Phys. J. C 45 (2006) 525-528.
\bibitem{4} D. Bohm, Quantum Mechanics, Printice Hall, Englewood Cliffs, NJ, 1951.
\bibitem{5} P. Senn, Threshold anomalies in one-dimensional scattering, Am. J. Phys. 56 (1988) 916-920.
\bibitem{6} M. S. de Bianchi, Levinson's theorem, zero-energy resonances, and time delay in one-dimensional scattering systems, J. Math. Phys. 35 (1994) 2719-2733.
\bibitem{7} N. Dombey, P. Kennedy, A. Calogeracos, Supercriticality and Transmission Resonances in the Dirac Equation, Phys. Rev. Lett. 85 (2000) 1787-1790.
\bibitem{8} P. Kennedy, N. Dombey, Low momentum scattering in the Dirac equation, J. Phys. A : Math. Gen. 35 (2002) 6645-6658.
\bibitem{9} P. Kennedy, The Woods-Saxon potential in the Dirac equation, J. Phys. A : Math. Gen. 35 (2002) 689-698.
\bibitem{10}  A. Calogeracos, N. Dombey, Klein tunnelling and the Klein paradox, Int. J. Mod. Phys. A 14 (1999) 631-644.
\bibitem{11} V. M. Villalba, C. Rojas, Scattering of a relativistic scalar particle by a cusp potential, Phys. Lett. A 362 (2007) 21-25.
\bibitem{12} V. M. Villalba, W. Greiner, Transmission resonances and supercritical states in a one-dimensional cusp potential, Phys. Rev. A 67 (2003) 052707 (4pp).
\bibitem{13} C. Rojas, V. M. Villalba, Scattering of a Klein-Gordon particle by a Woods-Saxon potential, Phys. Rev. A 71 (2005) 052101 (4pp).
\bibitem{14} V. M. Villalba, C. Rojas, Bound states of the Klein-Gordon equation in the presence of the short range potentials, Int. J. Mod. Phys. A 21 (2006) 313-326.
\bibitem{15} V. M. Villalba, L. A. Gonzalez-Arraga, Tunneling and transmission resonances of a Dirac particle by a double barrier, Phys. Scr. 81 (2010) 025010 (6pp).
\bibitem{16} K. Sogut, A. Havare, Transmission resonances in the Duffin–Kemmer–Petiau equation in (1+1)
dimensions for an asymmetric cusp potential, Phys. Scr. 82 (2010) 045013 (6pp).
\bibitem{17} A. Arda, O. Aydogdu, R. Sever, Scattering and bound state solutions of the asymmetric Hulth{\'e}n potential, Phys. Scr. 84 (2011) 025004 (6pp).
\bibitem{18} J. Y. Guo, X. Z. Fang, Scattering of a Klein–Gordon particle by a Hulthén potential, Can. J. Phys. 87 (2009) 1021-1024.
\bibitem{19} G. Bastard, Wave Mechanics Applied to Semiconductor Heterostructures, Les Ulis: Editions de Physique, 1988.
\bibitem{20} L. Serra, E. Lipparini, Spin response of unpolarized quantum dots, Europhys. Lett. 40 (1997) 667-672.
\bibitem{21} F. A. de Saavedra, J. Boronat, A. Pollas, A. Fabrocini, Effective mass of one 4He atom in liquid 3He, Phys. Rev. B 50 (1994) 4248-4251.
\bibitem{22} A. D. Alhaidari, Solution of the Dirac equation with position-dependent mass in the Coulomb field, Phys. Lett. A 322 (2004) 72-77.
\bibitem{23} A. D. Alhaidari, H. Bahlouli, A. Al Hasan, M. S. Abdelmonem, Relativistic scattering with a spatially
dependent effective mass in the Dirac equation, Phys. Rev. A 75 (2007) 062711 (14pp).
\bibitem{24} A. D. de Souza, C. Y. Jia, Classes of exact Klein–Gordon equations with spatially dependent masses: Regularizing
the one-dimensional inversely linear potential, Phys. Lett. A 352 (2006) 484-487.
\bibitem{25} L. Dekar, L. Chetouani, T. F. Hammann, An exactly soluble Schrödinger equation with smooth position-dependent mass, J. Math. Phys. 39 (1998) 2551-2563.
\bibitem{26} O. Panella, S. Biondini, A. Arda, New exact solution of the one-dimensional Dirac equation for the Woods–Saxon
potential within the effective mass case, J. Phys. A: Math. Theor. 43 (2010) 325302-(24pp).
\bibitem{27} L. Hulth{\'e}n, {\"U}ber die Eigenlösungen der Schrödingergleichung des Deuterons, Ark. Mat. Astron. Fys. 28 (1942) 1-12.
\bibitem{28} Y. P. Varshni, Eigenenergies and oscillator strengths for the Hulthen potential, Phys. Rev. A 41 (1990) 4682-4689.
\bibitem{29} M. Jameelt, Large N expansion for Hulthen potential, J. Phys. A: Math. Gen. 19 (1986) 1967-1972.
\bibitem{30} R. Barnana, R. Rajkumar, The shifted 1/N expansion and the energy eigenvalues of
the Hulthen potential for l $\neq$ 0, J. Phys. A: Math. Gen. 20 (1987) 3051-3056.
\bibitem{31} L. H. Richard, The Yukawa and Hulthen potentials in quantum mechanics, J. Phys. A: Math. Gen. 25 (1992) 1373-1382.
\bibitem{32} K. Sogut, A. Havare, Scattering of vector bosons by an asymmetric Hulthen potential, J. Phys. A: Math. Gen. 43 (2010) 225204 (14pp).
\bibitem{33} M. Abramowitz, I. A. Stegun, Handbook of Mathematical Functions, New York: Dover, 1970.
\bibitem{34} F. Dominguez-Adame, Bound states of the Klein-Gordon equation with vector and scalar Hulthen-type potentials, Phys. Lett. A 136 (1989) 175-177.
\bibitem{35} W. Greiner, B. Muller, J. Rafelski, Quantum Electrodynamics of Strong Fields, Springer-Verlag, Heidelberg, 1985.
\end{thebibliography}
\end{document}